\documentclass[runningheads, envcountsame, a4paper]{llncs}

\usepackage{multirow,bigstrut}
\usepackage{tabu}
\usepackage{wrapfig}
\usepackage{graphicx}
\usepackage{amsmath}
\usepackage{xcolor}
\usepackage[hyphens]{url}
\usepackage[misc]{ifsym}

\newcommand{\yt}{y_{t}}
\newcommand{\ythat}{\hat{y}_{t}}
\newcommand{\setT}{\mathcal{T}}

\newcommand{\setX}{\mathcal{X}}
\newcommand{\evect}{\mathbf{e}}

\newcommand{\setXstar}{\mathcal{X^*}}
\newcommand{\evectstar}{\mathbf{e^*}}
\newcommand{\setTstar}{\mathcal{T^*}}

\newcommand{\ses}{\texttt{ses}}
\newcommand{\holt}{\texttt{holt}}
\newcommand{\damped}{\texttt{damp}}
\newcommand{\mytheta}{\texttt{theta}}
\newcommand{\comb}{\texttt{comb}}
\newcommand{\rf}{\texttt{rf}}
\newcommand{\baseline}{\texttt{baseline}}
\newcommand{\arima}{\texttt{arima}}
\newcommand{\rwf}{\texttt{rwf}}
\newcommand{\tbats}{\texttt{tbats}}
\newcommand{\nn}{\texttt{nn}}
\newcommand{\tsensembler}{\texttt{ADE}}
\newcommand{\hyndmeta}{\texttt{FFORMS}}

\newcommand{\monitoring}{\textit{monitoring models}}
\newcommand{\monitored}{\textit{monitored models}}

\begin{document}
\title{Model Monitoring and Dynamic Model Selection in Travel Time-series Forecasting}
%
\titlerunning{Model Monitoring and Dynamic Model Selection in Time-series Forecasting}
\toctitle{Model Monitoring and Dynamic Model Selection in Travel Time-series Forecasting}
%
\author{Rosa Candela\inst{1} \and
Pietro Michiardi\inst{1} \and
Maurizio Filippone\inst{1} \and \\
Maria A. Zuluaga\inst{1,2}\Letter}
\authorrunning{R. Candela et al.}
\tocauthor{Rosa Candela {EURECOM}, Pietro Michiardi {EURECOM}, Maurizio Filippone {EURECOM}, Maria A. Zuluaga {EURECOM, Amadeus SAS}}
\institute{Data Science Department, EURECOM, Biot, France\\
\and
Amadeus SAS, Sophia Antipolis, France\\
\email{$\{$firstname.lastname$\}$@eurecom.fr}}

\maketitle              
\begin{abstract}
Accurate travel products price forecasting is a highly desired feature that allows customers to take informed decisions about purchases, and companies to build and offer attractive tour packages. Thanks to machine learning (ML), it is now relatively cheap to develop highly accurate statistical models for price time-series forecasting. However, once models are deployed in production, it is their monitoring, maintenance and improvement which carry most of the costs and difficulties over time. We introduce a data-driven framework to continuously monitor and maintain deployed time-series forecasting models' performance, to guarantee stable performance of travel products price forecasting models. Under a supervised learning approach, we predict the errors of time-series forecasting models over time, and use this predicted performance measure to achieve both model monitoring and maintenance. We validate the proposed method on a dataset of 18K time-series from flight and hotel prices collected over two years and on two public benchmarks.

\keywords{Model monitoring  \and Model maintenance \and Time-series \and Forecasting.}
\end{abstract}

\section{Introduction}

Travel industry actors, such as airlines and hotels, nowadays use sophisticated pricing models to maximize their revenue, which results in highly volatile fares~\cite{Chen2015}. For customers, price fluctuation  are a source of worry due to the uncertainty of future price evolution. This situation has opened the possibility to new businesses, such as travel meta-search engines or online travel agencies, providing decision-making tools to customers~\cite{Wohlfarth2011}. In this context, accurate price forecasting over time is a highly desired feature, as it allows customers to take informed decisions about purchases, and companies to build and offer attractive tour packages, while maximizing their revenue margin.

The exponential growth of computer power along with the availability of large datasets has led to a rapid progress in the machine learning (ML) field over the last decades. This has allowed the travel industry to benefit from the powerful ML machinery to develop and deploy accurate models for price time-series forecasting.  Development and deployment, however, only represent the first steps of a ML system's life cycle. Currently, it is the monitoring, maintenance and improvement of complex production-deployed ML systems which carry most of the costs and difficulties in time~\cite{Sculley2015,r2019overton}. Model monitoring refers to the task of constantly tracking a model's performance to determine when it degrades, becoming obsolete. Once a degradation in performance is detected, model maintenance and improvement take place to update the deployed model by rebuilding it, recalibrating it or, more generally, by doing model selection. 

While it is relatively easy and fast to develop ML-based methods for accurate price forecasting of different travel products, maintaining a good performance over time faces multiple challenges. Firstly, price forecasting of travel products involves the analysis of multiple time-series which are modeled independently, i.e.~a model per series rather than a single model for all. According to the 2019 World Air Transport Statistics report, almost 22K city pairs are directly connected by airlines through regular services~\cite{international2019world}. As each city pair is linked to a time-series, it is impossible to manually monitor the performance of every associated forecasting model. For scalability purposes, it is necessary to develop methods that can continuously and automatically monitor and maintain every deployed model. Secondly, time-series comprise time-evolving complex patterns, non-stationarities or, more generally, distribution changes over time, making forecasting models more prone to deteriorate over time~\cite{aiolfi_persistence_2006}. Poor estimations of a model's degrading performance can lead to business losses, if detected too late, or to unnecessary model updates incurring system maintenance costs~\cite{Sculley2015}, if detected too early. Efficient and timely ways to model monitoring are therefore key to continuously accurate in-production forecasts.
Finally, a model's degrading performance also implies that the model becomes obsolete. As a result, a specific model might not always be the right choice for a given series. Since time-series forecasting can be addressed through a large set of different approaches, the task of choosing the most suitable forecasting method requires finding systematic ways to carry out model selection efficiently. One of the most common ways to achieve all of this is cross-validation~\cite{arlot2010survey}. However, this approach is only valid at development and cannot be used to monitor and maintain models in-production due to the absence of ground truth data. 

In this work we introduce a data-driven framework to continuously monitor and maintain time-series forecasting models' performance in-production, i.e in the absence of ground truth, to guarantee continuous accurate performance of travel products price forecasting models.
Under a supervised learning approach, we predict the forecasting error of time-series forecasting models over time. We hypothesize that the estimated forecasting error represents a surrogate measure of the model's future performance. As such, we achieve continuous monitoring by using the predicted forecasting error as a measure to detect degrading performance. Simultaneously, the predicted forecasting error enables model maintenance by allowing to rank multiple models based on their predicted performance, i.e. model comparison,  and then select the one with the lowest predicted error measure, i.e. model selection. We refer to it as a model monitoring and model selection framework.

The remaining of this paper is organized as follows.
Section~\ref{sect_rw} discusses related work. 
Section~\ref{sect_background} reviews the fundamentals of time-series forecasting and performance assessment. 
Section~\ref{sect_methodology} describes the proposed model monitoring and maintenance framework. 
Section~\ref{sect_experiments} describes our datasets and presents the experimental setup. Experiments and results are discussed in section~\ref{sec:results}.  Finally, in section~\ref{sect_conclusions} we summarize our work and discuss key findings.

\section{Related Work} \label{sect_rw}
\textbf{Maintainable industrial ML systems.} Recent works from tech companies~\cite{Baylor2017,lin2012large,r2019overton} have discussed their strategies to deal with some of the so-called \textit{technical debts}~\cite{Sculley2015} in which ML systems can incur when in production. These works mainly focus on the hard- and soft-ware infrastructure used to mitigate these \textit{debts}. Less emphasis is given to the specific methods put in place.
\\
\textbf{Concept drift.} The phenomenon of time-evolving data patterns is known as concept drift. As time-series are not strictly stationary, it is a common problem of time-series forecasting usually addressed through regular model updates. Most works have focused on its detection, what we denote model monitoring, without performing model selection as they are typically limited to a single model~\cite{Ferreira:2014:DCT:2542820.2562373,10.1007/978-3-642-34166-3_40}. The exception to this is the work of \cite{saadallah2019drift,Saadallah2020}, where a weighted sliding-window is used to combine the forecasts of multiple candidate models into a single value.
\\
\textbf{Performance assessment without ground truth.}
An alternative to cross-validation is represented by information criteria. The rationale consists in quantifying the best trade-off between models' goodness of fit and simplicity. 
Information criteria are mostly used to compare nested models, whereas the comparison of different models requires to compute likelihoods on the same data. 
Being fully data-driven, our framework avoids any constraint regarding the candidate models, leading to a more general way to perform model selection.
Specifically to time-series forecasting, Wagenmakers et al.~\cite{wagenmakers2006accumulative} achieve performance assessment in the absence of ground truth using a concept similar to ours. They estimate the forecasting error of a new single data point by adding previously estimated forecast errors, obtained from already observed data points. The use of the previous errors makes it sensible to unexpected outlier behaviors of the time-series.
\\
\textbf{Meta-learning.} Meta-learning has been proposed as a way to automatically perform model selection. Its performance has been recently demonstrated in the context of time-series forecasting. Both~\cite{ALI20189,RePEc:msh:ebswps:2018-6} formulate the problem as a supervised learning one, where the meta-learner receives a time-series and outputs the ``best'' forecasting model. Authors in~\cite{cerqueira2017arbitrated} share our idea that forecasting performance decays in time, thus they train a meta-learner to model the error incurred by the base models at each prediction step as a function of the time-series features. Differently from~\cite{RePEc:msh:ebswps:2018-6}, our approach does not seek to select a different model family for each time-series, and avoids model selection at each time step~\cite{cerqueira2017arbitrated}, since these two represent expensive overheads for in-production maintenance. Instead, we maintain a fast forecasting procedure and select the best model for a given time period in the future, which length can be relatively high (6-9 months, for instance).

\section{Time-series forecasting and performance measures} \label{sect_background}

A univariate time-series is a series of data points 
$\smash{\setT = \{ y_1, \ldots, y_T \}}$,
each one being an observation of a process measured at a specific time $t$. 
Univariate time-series contain a single variable at each time instant, while multivariate time-series record more than one variable at a time. Our application is concerned with univariate time-series, which are recorded at discrete points in time, e.g., monthly, daily, hourly. However, extension to the multivariate setting is straightforward. 

Time-series forecasting is the task consisting in the use of these past observations (or a subset thereof) to predict future values $\setT_h = \{ \hat{y}_{T+1}, \ldots, \hat{y}_{T+h}\}$, with $h$ indicating the forecasting horizon. The number of well-established methods to perform time-series forecasting is quite large. Methods go from classical statistical methods, such as Autoregressive Moving Average (ARMA) and Exponential smoothing, to more recent machine learning models which have shown  outstanding performance in different tasks, including time-series forecasting.

The performance assessment of forecasting methods is commonly done using error measures. Despite decades of research on the topic, there is still not an unanimous consensus on the best error measure to use among the multiple available options~\cite{HYNDMAN2006679}. 
Among the most used ones, we find Symmetric Mean Absolute Percentage Error (sMAPE) and Mean Absolute Scaled Error (MASE). These two have been adopted in recent time-series forecasting competitions~\cite{article}. 

\section{Monitoring and model selection framework} \label{sect_methodology}
\begin{figure}[t]
	\begin{center}
		\includegraphics[width=0.8\textwidth]{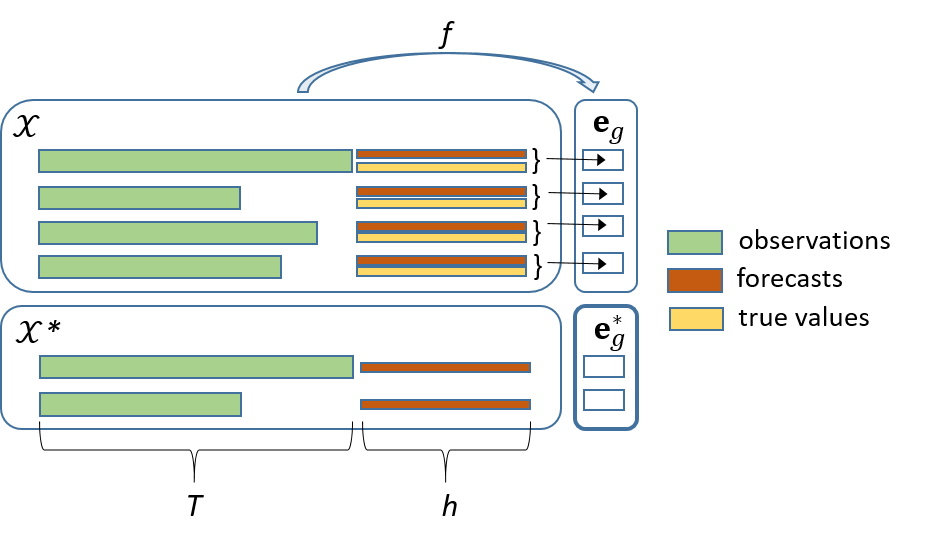}
		\caption{Illustration of the proposed method. $\setX{}$ and $\setXstar$ contain multiple time-series, each of these composed of $T_i$ observations (green) and $h$ forecasts (red) estimated by a \textit{monitored model}, $g$. $\evect_g$ represents the forecasting performance of the \textit{monitored model}. It is computed using the true values (yellow). A \textit{monitoring model} is trained to learn the function $f$ mapping $\setX$ to $\evect_g$. With the learned $f$, the \textit{monitoring model} is able to predict $\evect^{*}_{g}$, the predicted forecasting performance of the  \textit{monitored model} given $\setXstar$. 
		} \label{fig:fig_supervised}
	\end{center}
\end{figure}

Let us denote 
$\setX=\smash{\{ \setT^{(i)}, \setT_{h}^{(i)} \}_{i=1}^{N}}$
the input training set. A given input $i$ is formed by the observed time-series $\smash{\setT^{(i)}}$ and $h$ forecasted values, $\smash{\setT_{h}^{(i)}}$. The values in $\smash{\setT_{h}^{(i)}}$ are obtained by a given forecasting model which we hereby denote a \textit{monitored model}, $g$. Let $\evect_{g}=\smash{ \{ e^{(i)}_g  \}_{i=1}^{N}}$ be a collection of $N$ performance measures assessing the accuracy of the forecasts $\smash{\setT_{h}^{(i)}}$ estimated by $g$. A given performance measure $\smash{e^{(i)}_g}$ is obtained by comparing the forecasts $\smash{\setT_{h}^{(i)}}$ from $g$ to the true values. 

Lets define a \textit{monitoring model} as a model that is trained to learn a function $f$ mapping the input time-series $\setX$ to the target $\smash{\evect_g}$. 
Given a new set of time-series 
$\setXstar=\smash{\{ \setTstar^{(i)}, \setTstar_{h}^{(i)}  \}_{i=1}^{N^*}}$, 
formed by a time-series of observations $\smash{\setTstar^{(i)}}$, $\smash{|\setTstar^{(i)}|=T^*_i}$,  and $h$ forecasts $\smash{\setTstar_{h}^{(i)}}$ obtained by $g$, the learned \textit{monitoring model} predicts $\evect^{*}_g$, i.e. the predicted performance measure of $g$ given $\setXstar$ (Figure~\ref{fig:fig_supervised}).

The predicted performance measures $\smash{\evect^{*}_g}$ represent a surrogate measure of the performance of a given $g$ within the forecasting horizon $h$. As such it is used for the two tasks: model monitoring and selection. Model monitoring is achieved by using $\smash{\evect^{*}_g}$ as an alert signal. If the estimated performance measure of the \textit{monitored model} is poor, this means the model has become stale. To achieve model selection, $\smash{\evectstar}$ are used to rank multiple \monitored{} and choose the one with the best performance 
If the two tasks are executed in a continuous fashion over time, it is possible to guarantee accurate forecasts in an automated way. 
 
In the following, we describe the performance measure $\textbf{e}$ that we use in our framework, as well as the \textit{monitoring} and \textit{monitored models} that we chose to validate our hypotheses.

\subsection{Performance measure}
As previously discussed, performance accuracy of time-series forecasts is measured using error metrics. In this framework, we use the sMAPE. It is defined as:
%
\begin{equation} \label{sMAPE_Eq}
\textrm{sMAPE} = \frac{1}{h}\sum_{t=1}^{h}2\frac{\left | \yt - \ythat \right |}{\left |\yt \right |+\left | \ythat \right |} \text{,}
\end{equation}
where $h$ is the number of forecasts (i.e. forecasting horizon), $y$ is the true value and $\hat{y}$ is the forecast. 

In the literature, there are multiple definitions of the sMAPE. We choose the one introduced in~\cite{chen2004assessing} because it is bounded between 0 and 2; specifically, it has a maximum value of 2 when either $y$ or $\hat{y}$  is zero, and it is zero when the two values are identical. The sMAPE has two important drawbacks: it is undefined when both $y$, $\hat{y}$ are zero and it can be numerically unstable when the denominator in Eq.~\ref{sMAPE_Eq} is close to zero. In the context of our application, this is not a problem since it is unlikely to have prices with value zero or very close to it.

%
%
%
\subsection{Monitoring models} \label{sect_models}
The formulation of our framework is generic in the sense that any supervised technique that can solve regression problems can be used as a \textit{monitoring model}. In this work, we decided to focus on latest advances in deep learning. We consider four alternative \monitoring{}: Long Short-Term Memory (LSTM) networks, Convolutional Neural Networks (CNNs), Bayesian CNNs and Gaussian processes (GP). The latter two models differ from the former ones in that they also provide uncertainties around the  predictions. This can enrich the output provided by the monitoring framework, in that whenever an alert is issued because of poor performance, this is equipped with information about its reliability This section illustrates the basic ideas of each of the selected \monitoring{}.
\\
\\
\textbf{Long Short-Term Memory networks.}
LSTM~\cite{hochreiter1997long} networks are a type of Recurrent Neural Networks (RNNs)  that solve the issue of the vanishing gradient problem~\cite{bengio1994learning} present in the original RNN formulation. They achieve this by introducing a cell state into each hidden unit, which memorizes information. As RNNs they are a well-established architecture 
to model sequential data. By construction, LSTMs can handle sequences of varying length, with no need for extra processing like padding. This is useful in our application, whereby time-series in the datasets have different lengths.
%

\noindent
\textbf{Convolutional Neural Networks.}
CNNs~\cite{lecun1998gradient} are particular class of deep neural networks where the weights (filters) are designed to promote local information to propagate from the input to the output at increasing levels of granularity.  
We use the original LeNet \cite{LeCun:1999:ORG:646469.691875} architecture, as it obtains generally good results in image recognition problems, while being considerably faster to train with respect to more modern architectures. 
CNNs are not originally conceived to work with time-series data. We adapt the architecture to work with time-series by using 1D convolutional filters. 
Unlike RNNs, this model does not support inputs of variable size, so we to resort to padding: where necessary we append zeros to a time-series to make them uniform in length. We denote this model LeNet.

\noindent
\textbf{Bayesian Convolutional Neural Networks.}
Bayesian CNNs~\cite{gal2016dropout} represent the probabilistic version of CNNs, used in applications where quantification of the uncertainty in predictions is needed. 
Network parameters are assigned a prior distribution and then inferred using Bayesian inference techniques.
Due to the intractability of the problem, the use of approximations is required.
Here we choose Monte Carlo Dropout~\cite{gal2016dropout} as a practical way to carry out approximate Bayesian CNNs. By applying dropout at test time we are able to sample from an approximate posterior distribution over the network weights. We use this technique on the LeNet CNN with 1D filters to produce probabilistic outputs. We denote this model Bayes-LeNet.

\noindent
\textbf{Gaussian processes.}
GPs~\cite{Rasmussen:2005:GPM:1162254} form the basis of probabilistic nonparametric models. Given a supervised learning problem, GPs consider an infinite set of functions mapping input to output. These functions are defined as random variables with a joint Gaussian distribution, specified by a mean function and a covariance function, the latter encoding properties of the functions with respect to the input. 
One of the strengths of GP models is the ability to characterize uncertainty regardless of the size of the data. Similarly to CNNs, in this model input sequences must have the same length, so we resort to padding.

%
%
\subsection{Monitored models}
Similar to \monitoring{}, given the generic nature of the proposed framework, there is no constraint on the type of \monitored{} that can be used. Any time-series forecasting method can be monitored. For this proof of concept, we consider six different \monitored{}. We select five of them from the ten benchmarks provided in the M4 competition~\cite{article}, a recent time-series forecasting challenge. These are: Simple Exponential Smoothing (\ses), Holt's Exponential Smoothing (\holt), Dampen Exponential Smoothing (\damped), Theta (\mytheta) and a combination of  \ses{} - \holt{} - \damped{} (\comb).  
Besides these five methods, we included a simple Random Forest (\rf) 
, in order to enrich the benchmark with a machine learning-based model. We refer the reader to \cite{breiman2001random,article} for further details on each of these approaches.

%
%

%
\section{Experimental setup} \label{sect_experiments}
This section presents the data, provides details about the implementation of our methods to ease reproducibility and concludes by describing the evaluation protocol carried during the experiments.

\subsection{Data}
\textbf{Flights and hotels datasets.} We focus on two travel products: direct flights between city pairs and hotels. Our data is an extract of prices for these two travel products obtained from the Amadeus for Developers API~\footnote[1]{\url{https://developers.amadeus.com/}}, an online web-service which enables access to travel-related data through several APIs. It was collected over a two-years and one-month period. Table~\ref{tab_datasets} presents some descriptive features of the datasets.

Using the service's Flight Low-fare Search API, we collected daily data for one-way flight prices of the top 15K most popular city pairs worldwide. The collection was done in two stages. A first batch, corresponding to the top 1.4K pairs (\textsc{flights}), was gathered for the whole collection period. The second batch, corresponding to the remaining pairs (\textsc{flights-ext}), was collected only over the second year. For hotels, we used the Hotel API to collect daily hotel prices for a two-night stay at every destination city contained in the top city pairs used for flight search. These represent 3.2K different time-series.  

Both APIs provide information about the available offers for flights/hotels, that meet the search criteria (origin-destination and date, for flights; city, date and number of nights, fixed to 2, for hotels) at the time of search. As such, it is possible to have multiple offers (flights or hotel rooms) for a given search criteria. When multiple offers were proposed, we averaged the different prices to have a daily average flight price for a given city pair, in the case of flights, or daily average hotel price for a given city, in the case of hotels. In the same way, it is possible to have no offers for a given search criteria. Days with no available offers were reported as missing data. Lack of offers can be caused by sold outs, specific flight schedules (e.g. no daily flights for a city pair) or seasonal patterns (e.g. flights for a part of the year or seasonal hotel closures). More rarely, they could even be due to a failure in the query sent to the API. As a result, the number of available observations is smaller than the length of the collection period (see Table~\ref{tab_datasets}).   
\\
\\
\noindent
\textbf{Public benchmarks.} In addition to travel products data, we decided to include data coming from publicly available benchmarks. Benchmark data are typically curated and avoid problems present in real data, such as those previously discussed regarding missing data, allowing for an objective assessment and more controlled setup for experimentation. We included two sets from the M4 time-series forecasting challenge competition~\cite{article} dataset, \textsc{yearly} and \textsc{weekly}. Table~\ref{tab_datasets} presents statistics on the number of time-series and the available number observations per time-series for these two datasets. Here, the number of available observations is equivalent to the time-series length as no time-series contains missing values. 
 
\begin{table}[t]
	\centering
	\caption{Information about number of time-series, and  minimum (min-obs), maximum (max-obs), mean (mean-obs) and standard deviation (std-obs) of the available number of time-series observations per dataset.}
	\label{tab_datasets}
	\setlength{\tabcolsep}{0.23em}
	\begin{tabular}{cccccc} 
		\hline
		\textbf{Name}  & \multicolumn{1}{l}{\textbf{\# time-series} } & \multicolumn{1}{l}{\textbf{min-obs} } & \multicolumn{1}{l}{\textbf{max-obs} } & \multicolumn{1}{l}{\textbf{mean-obs}} & \multicolumn{1}{l}{\textbf{std-obs} }  \\ 
		\hline
		\textsc{flights}         & 1,415                                        & 431                                      & 745                                      & 734                                   & 23                                     \\
		\textsc{flights-ext}     & 13,810                                       & 50                                       & 347                                      & 346                                   & 13                                     \\
		\textsc{hotels}          & 3,207                                        & 1                                        & 658                                      & 368                                   & 128                                    \\
		\textsc{yearly}     & 23,000                                       & 13                                       & 835                                      & 31                                   & 25                                     \\
		\textsc{weekly}     & 359                                       & 80                                       & 2,597                                      & 1022                                   & 706                                     \\
		\hline
	\end{tabular}
\end{table}

\subsection{Implementation}
The LSTM network was implemented in Tensorflow. It is composed of one hidden layer with 32 hidden nodes. It is a dynamic LSTM, in that it allows the input sequences to have variable lengths, by dynamically creating the graph during execution.
The two CNN-based \monitoring{} use the LeNet architecture. We modified both convolutional and pooling layers with 1D filters, given that the input of the model consists in sequences of one dimension. We added dropout layers to limit overfitting. In the Bayesian CNN, we applied a dropout rate of 0.5, also at testing time, to obtain 100 Monte Carlo samples as approximation of the true posterior distribution. 
The GP model used the implementation of Sparse GP Regression from the GPy library\footnote[2]{\url{http://github.com/SheffieldML/GPy}}. The inducing points~\cite{titsias2009variational} were initialized with $K$-means and were then fixed during optimization. We used a variable number of inducing points depending on the size of the input and a RBF kernel with Automatic Relevance Determination (ARD). In all experiments we used 75\% data for training and 25\% for test and the Adam optimizer with default learning rate~\cite{DBLP:journals/corr/KingmaB14}. Only in the dataset \textsc{flights-ext} we used mini-batches of size $128$ to speed up the training.
%
For the \monitored{}, we used the implementation available from the M4 competition benchmark Github repository\footnote[3]{\url{https://github.com/M4Competition/M4-methods}} and 
we used the Python sklearn package~\cite{scikit-learn} implementation of R   andom Forest. All code has been made publicly available\footnote[4]{\url{https://github.com/robustml-eurecom/model_monitoring_selection}}.

\subsection{Evaluation protocol}
For flight and hotel data we set $h=\{90,180\}$, which means we are predicting the price for $h$ days ahead. These are two commonly used values in travel, representing 3 and 6 months ahead of the planned trip, so it is important to have accurate predictions over those horizons. For the M4 competition datasets, we use the horizon given by the challenge organizers: $h=6$ for \textsc{yearly} and $h=13$ for \textsc{weekly}.
For each dataset, we reserve the first $T_i$ data points of the \textit{i-th} time-series, where $T_i$ depends on the time-series's length, as input of the \monitored{} to obtain $h$ forecasts. Where missing values were found, in flights or hotels, these were replaced with the nearest non-missing value in the past. We build $\setX$ and $\setXstar$, by taking 75\% and 25\% from the total number of time-series, respectively. We thus compute the forecasting errors $\evect$ using the sMAPE in Eq. \ref{sMAPE_Eq} for the training set $\setX$. Finally, we predict the performance measure $\evectstar$ for the time-series in $\setXstar$, using the four \monitoring{}. 

We compare our model monitoring and selection framework  with the standard cross-validation method, which we here denote \baseline, where a model's estimated performance is obtained ``offline'' at training time with the available data. Specifically, given $T$ observations, we use the last $h$ observations as validation set to evaluate the model. This implies to reduce the number of observations available to train the forecasting models, which can be problematic when either $T$ is small or $h$ is large. 
\section{Experiments and Results} \label{sec:results}

We first study the proposed framework's ability to achieve model monitoring (Sec~\ref{sec:monitorperf}). Then, we demonstrate how the predicted forecasting errors can be used to carry out model selection and how it positions w.r.t state-of-the-art methods doing the same task (Sec~\ref{sec:selperf}). In Sec~\ref{sec:allperf}, we illustrate the performance of the joint model monitoring and selection framework in our target application.

\subsection{Model monitoring performance}\label{sec:monitorperf}
\begin{wrapfigure}{L}{0.6\textwidth}
	\centering
	\includegraphics[width=0.59\textwidth]{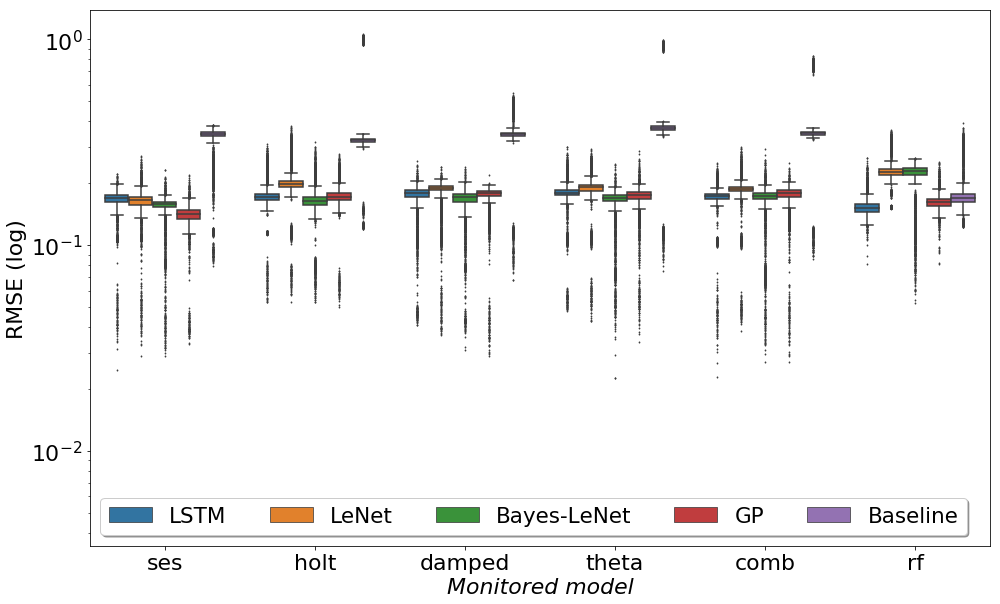}
	\caption{RMSE between predicted and measured forecasting error (sMAPE) on all datasets (log scale). The reported baseline RMSE is obtained by comparing the estimated sMAPE at training with the observed values at testing.}
	\label{fig:rmse_summary}
\end{wrapfigure}

We evaluate if the \monitoring{}' predicted sMAPEs can be used for model monitoring by estimating if the predicted measure represents a good estimate of a \textit{monitored model}'s future forecasting performance. We assess the quality of the predicted forecasting errors by estimating the root mean squared error (RMSE) between the predicted sMAPEs and the true sMAPEs, for every \textit{monitored model}. The true sMAPE is obtained using the \textit{monitored model}'s predictions and the time-series' observations in through Eq.~\ref{sMAPE_Eq}. As a reference, we report the \baseline{} RMSE, which is obtained by comparing the estimated sMAPE at training with the observed values at testing.Figure \ref{fig:rmse_summary} left summarizes the obtained results on all datasets. 

The overall average error incurred by the \monitoring{} is low. This suggests that the forecasting error predictions are accurate, meaning that it is reliable to carry out model monitoring.  When compared to it, the \monitoring{} consistently perform better than standard cross-validation when estimating the future performance of the forecasting \monitored{}. There is an exception to this when the \textit{monitored model} is the Random Forests (\rf{}). In this case, the \baseline{} is not the worst performing approach. However, it is still surpassed in performance by both LSTM and GP.

\begin{figure}[t]
	\centering
		\includegraphics[width=\textwidth]{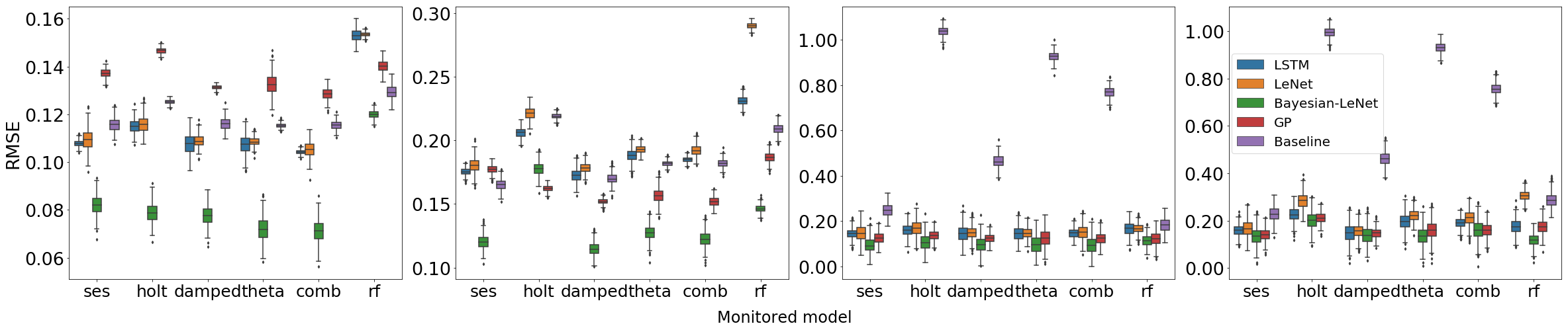}
	\caption{RMSE between predicted and measured forecasting error (sMAPE). From left to right \textsc{flights} and \textsc{flights-ext} (top) with 1) $h=90$, 2) $h=180$, hotels 3) $h=90$, 4) $h=180$.
	}	\label{fig:fig_rmse_results}
\end{figure}

Figure~\ref{fig:fig_rmse_results} details the  results obtained for flights and hotels time-series. Table~\ref{tab:summary-travel} stratifies the results for travel product time-series in terms of the forecasting horizon. results show that Bayes-LeNet obtains the lowest RMSEs, whereas GPs follows closely and reports lower standard deviation.
Overall, our approach outperforms the \baseline{} for large forecasting horizons, e.g. $h=180$, while the methods get closer as the forecasting horizon decreases. This is consistent with our hypothesis that data properties change over time. Using a validation set composed of time points close to the unseen data gives consistent information about the model's performance, because the two sets of data (validation and unseen data) have similar properties. However, increasing $h$ has the effect of pushing away the validation time points from the unseen data. In this case, it is better to rely on the forecast error prediction rather than on an error measure obtained during training. 

\begin{table}[t]
	\centering
	\caption{RMSE between predicted and true sMAPEs for flights and hotel time-series.}\label{tab:summary-travel}
	\setlength{\tabcolsep}{0.23em}
	\begin{tabular}{|l|r|r|r|r|}
		\hline
		\multicolumn{1}{|c|}{\textbf{Monitoring}} &  \multicolumn{2}{c|}{\textbf{Flights}} & \multicolumn{2}{c|}{\textbf{Hotels}}\\
		\cline{2-5}
		\multicolumn{1}{|c|}{\textbf{model}}& \multicolumn{1}{c|}{$h=90$} &  \multicolumn{1}{c|}{$h=180$} & \multicolumn{1}{c|}{$h=90$} &  \multicolumn{1}{c|}{$h=180$}\\
		\hline
		LSTM   & 0.116 $\pm$ 0.017 & 0.151 $\pm$ 0.031& 0.193  $\pm$ 0.021  & 0.182 $\pm$ 0.039\\
		LeNet  & 0.117 $\pm$ 0.017& 0.155 $\pm$ 0.031 & 0.209  $\pm$ 0.039& 0.224  $\pm$ 0.062\\
		Bayes-LeNet &\textbf{ 0.084 $\pm$ 0.017}  & \textbf{0.100 $\pm$ 0.035} & \textbf{0.135  $\pm$ 0.022} &\textbf{ 0.148  $\pm$ 0.044}\\
		GP & 0.136 $\pm$ 0.007 & 0.126 $\pm$ 0.028 & 0.164  $\pm$ 0.014 & 0.165  $\pm$ 0.036\\
		\hline
		\baseline{} & 0.119 $\pm$ 0.006 & 0.604 $\pm$ 0.328 & 0.190  $\pm$ 0.020 & 0.609  $\pm$ 0.302\\
		\hline
	\end{tabular}
\end{table}

\subsection{Model selection performance}\label{sec:selperf}
In this experiment, 
we assess the capacity of the proposed method to assist model selection in the absence of ground truth. \textit{Monitored models} are ranked by estimating the average predicted sMAPE over a given time-series and ordering the resulting values in ascending order. In this way, we obtain a list of \monitored{} from the best to the worst one. The best performing \textit{monitored model} is selected.
We compare the ground truth ranking with the one obtained by each of the \monitoring{} and the \baseline{}. We apply a Wilcoxon test \cite{wilcoxon1945individual} to the ranking results to verify if there are significant differences between each of the ranked \monitored. Table \ref{rank_table} presents obtained results in hotels and flights. 
%
%

\begin{table}[t]
	\centering
	\caption{Comparison between true and predicted model rankings, in ascending order of sMAPE. Underlined values indicate pairs of forecasting models not significantly different, according to Wilcoxon test.}
	\resizebox{0.99\textwidth}{!}{
		\label{rank_table}
		\begin{tabular}{|clr|lr|lr|lr|lr|lr|} 
			\hline
			\multicolumn{1}{|l}{}            & \multicolumn{2}{c|}{\multirow{2}{*}{ \textbf{Ground Truth} }} & \multicolumn{8}{c|}{\textbf{Monitoring models} }                                                                                                                & \multicolumn{2}{c|}{\multirow{2}{*}{ \textbf{Baseline} }}          \\ 
			\cline{4-11}
			& \multicolumn{2}{c|}{}                                         & \multicolumn{2}{c|}{\textbf{LSTM} }                 & \multicolumn{2}{c|}{\textbf{LeNet} }          &
			\multicolumn{2}{c|}{\textbf{Bayes-LeNet} }          & \multicolumn{2}{c|}{\textbf{GPs} }                  & \multicolumn{2}{c|}{}                                              \\ 
			\hline
			\multicolumn{13}{c}{\textsc{hotels} - $h$ = 180 }                                                                                                                                                                                                                                                                                \\ 
			\hline
			& \multicolumn{1}{c}{\textbf{model}}  & \multicolumn{1}{c|}{\textbf{sMAPE}}           & \multicolumn{1}{c}{\textbf{model}}  & \multicolumn{1}{c|}{\textbf{sMAPE}} & \multicolumn{1}{c}{\textbf{model}}  & \multicolumn{1}{c|}{\textbf{sMAPE}}   & \multicolumn{1}{c}{\textbf{model}}  & \multicolumn{1}{c|}{\textbf{sMAPE}}          & \multicolumn{1}{c}{\textbf{model}}   & \multicolumn{1}{c|}{\textbf{sMAPE}}  & \multicolumn{1}{c}{\textbf{model}}  & \multicolumn{1}{l|}{\textbf{sMAPE} }  \\
			\hline
			\textbf{1}                       & \damped           & 0.244 (0.153) & \ses & 208 (0.015)             & \ses              & 0.208 (0.032)    & \ses             & 0.212 (0.087)           & \damped           & \underline{0.230} (0.119)   & \ses              & 0.326 (0.202)                        \\
			\textbf{2}                       & \ses              & 0.246 (0.164)            & \damped           & 0.220 (0.033) & \damped & 0.211 (0.056)   & \damped          & 0.224 (0.130)           & \ses              & \underline{0.231} (0.121)   & \rf               & \underline{0.413} (0.333)                        \\
			\textbf{3}                       & \mytheta            & 0.269 (0.217)            & \mytheta            & 0.233 (0.059) & \mytheta & \underline{0.231} (0.024)    & \comb            & \underline{0.249} (0.166)           & \comb             & 0.251 (0.149)   & \damped           & \underline{0.462} (0.391)                        \\
			\textbf{4}                       & \comb             & 0.270 (0.207)            & \comb             & 0.234 (0.057) & \rf & \underline{0.236} (0.047)    & \mytheta           & \underline{0.268} (0.234)           & \mytheta            & 0.252 (0.160)   & \comb             & 0.746 (0.569)                        \\
			\textbf{5}                       & \rf               & 0.316 (0.300)            & \holt             & 0.280 (0.124) & \comb & 0.278 (0.145)    & \rf              & 0.324 (0.329)           & \rf               & 0.291 (0.207)   & \mytheta            & 0.938 (0.620)                        \\
			\textbf{6}                       & \holt             & 0.325 (0.277)            & \rf               & 0.292 (0.210) & \holt & 0.298 (0.189)   & \holt            & 0.325 (0.162)           & \holt             & 0.299 (0.190)   & \holt             & 1.047 (0.660)                        \\ 
			\hline
			\multicolumn{13}{c}{\textsc{hotels} - $h$ = 90 }                                                                                                                                                                              \\ 
			\hline
			& \multicolumn{1}{c}{\textbf{model}}  & \multicolumn{1}{c|}{\textbf{sMAPE}}           & \multicolumn{1}{c}{\textbf{model}}  & \multicolumn{1}{c|}{\textbf{sMAPE}} & \multicolumn{1}{c}{\textbf{model}}  & \multicolumn{1}{c|}{\textbf{sMAPE}}   & \multicolumn{1}{c}{\textbf{model}}  & \multicolumn{1}{c|}{\textbf{sMAPE}}          & \multicolumn{1}{c}{\textbf{model}}   & \multicolumn{1}{c|}{\textbf{sMAPE}}  & \multicolumn{1}{c}{\textbf{model}}  & \multicolumn{1}{l|}{\textbf{sMAPE} }  \\
			\hline
			\textbf{1}                       & \damped           & \underline{0.242} (0.175)            & \ses              & 0.203 (0.022) & \ses & 0.217 (0.065)    & \ses             & 0.238 (0.088)           & \damped           & 0.221 (0.137)   & \comb             & 0.237 (0.166)                        \\
			\textbf{2}    & \ses              & \underline{0.243} (0.174)            & \damped           & 0.218 (0.026) & \damped & 0.223 (0.073)    & \damped          & 0.239 (0.122)           & \comb             & 0.238 (0.155)   & \ses              & \underline{0.239} (0.177)                        \\
			\textbf{3} & \comb             & \underline{0.253} (0.189)            & \mytheta            & 0.223 (0.022) & \mytheta & 0.227 (0.063)    & \comb            & 0.259 (0.108)           & \mytheta            & 0.240 (0.151)   & \damped           & \underline{0.250} (0.194)                        \\
			\textbf{4} & \mytheta            & 0.254 (0.190)            & \comb             & 0.224 (0.030) & \comb & 0.229 (0.047)   & \mytheta           & 0.263 (0.132)           & \ses              & 0.244 (0.180)   & \mytheta            & 0.251 (0.201)                        \\
			\textbf{5} & \holt             & 0.275 (0.217)            & \holt             & 0.244 (0.052) & \holt & 0.252 (0.096)   & \holt            & 0.282 (0.190)           & \holt             & \underline{0.265} (0.185)   & \holt             & 0.277 (0.235)                        \\
			\textbf{6} & \rf               & 0.293 (0.285)            & \rf               & 0.254 (0.103) & \rf & 0.263 (0.059)   & \rf              & 0.298 (0.176)           & \rf               & \underline{0.266} (0.191)   & \rf               & 0.311 (0.296)                        \\ 
			\hline
			\multicolumn{13}{c}{\textsc{flights} - $h$ = 180 }                                                                                                                                                                                                                \\ 
			\hline
			& \multicolumn{1}{c}{\textbf{model}}  & \multicolumn{1}{c|}{\textbf{sMAPE}}           & \multicolumn{1}{c}{\textbf{model}}  & \multicolumn{1}{c|}{\textbf{sMAPE}}   & \multicolumn{1}{c}{\textbf{model}} & \multicolumn{1}{c|}{\textbf{sMAPE}}   & \multicolumn{1}{c}{\textbf{model}}  & \multicolumn{1}{c|}{\textbf{sMAPE}}       & \multicolumn{1}{c}{\textbf{model}}  &\multicolumn{1}{c|}{\textbf{sMAPE}}  & \multicolumn{1}{c}{\textbf{model}}  & \multicolumn{1}{c|}{\textbf{sMAPE}}                       \\
			\hline
			\textbf{1}                       & \rf             & 0.238 (0.163)            & \rf             & 0.203 (0.007) & \rf & 0.199 (0.039)   & \rf            & 0.219 (0.075)           & \rf           & 0.213 (0.108)   & \rf              & 0.259 (0.200)                        \\
			\textbf{2}                       & \ses           & 0.247 (0.144)            & \mytheta           & \underline{0.217} (0.026) & \ses & 0.215 (0.012)   & \mytheta          & 0.220 (0.097)           & \mytheta             & \underline{0.226} (0.098)   & \damped            & \underline{0.277} (0.150)                        \\
			\textbf{3}                       & \mytheta            & 0.248 (0.175)            & \ses            & \underline{0.218} (0.024) & \damped & \underline{0.216} (0.074)    & \ses           & 0.233 (0.098)           & \ses            & \underline{0.227} (0.100)   & \ses           & \underline{0.278} (0.151)                        \\
			\textbf{4}                       & \damped              & 0.249 (0.144)            & \damped              & 0.219 (0.022) & \mytheta & \underline{0.217} (0.042)   & \damped           & 0.240 (0.090)           & \damped              & 0.229 (0.098)   & \mytheta             & 0.281 (0.155)                        \\
			\textbf{5}                       & \comb             & 0.250 (0.148)            & \comb             & 0.221 (0.027) & \comb & 0.219 (0.016)   & \comb            & 0.241 (0.094)           & \comb             & 0.231 (0.054)   & \comb             & 0.283 (0.160)                        \\
			\textbf{6} & \holt               & 0.260 (0.162)            & \holt               & 0.223 (0.034) & \holt & 0.222 (0.036)   & \holt              & 0.250 (0.088)           & \holt               & 0.238 (0.119)   & \holt               & 0.299 (0.199)  \\
			\hline
			\multicolumn{13}{c}{\textsc{flights} - $h$ = 90 }                                                                                                                                                                                                                \\ 
			\hline
			& \multicolumn{1}{c}{\textbf{model}}  & \multicolumn{1}{c|}{\textbf{sMAPE}}           & \multicolumn{1}{c}{\textbf{model}}  & \multicolumn{1}{c|}{\textbf{sMAPE}}   & \multicolumn{1}{c}{\textbf{model}} & \multicolumn{1}{c|}{\textbf{sMAPE}}   & \multicolumn{1}{c}{\textbf{model}}  & \multicolumn{1}{c|}{\textbf{sMAPE}}       & \multicolumn{1}{c}{\textbf{model}}  & \multicolumn{1}{c|}{\textbf{sMAPE}}  & \multicolumn{1}{c}{\textbf{model}}  & \multicolumn{1}{c|}{\textbf{sMAPE}}                       \\
			\hline
			\textbf{1}                       & \comb             & 0.174 (0.102)            & \comb             & 0.154 (0.081) & \mytheta & 0.160 (0.067)   & \comb            & 0.151 (0.086)           & \damped           & 0.159 (0.073)   & \ses              & \underline{0.187} (0.110)                        \\
			\textbf{2}                       & \damped           & 0.175 (0.106)            & \damped           & 0.155 (0.076) & \comb & 0.164 (0.085)   & \damped          & 0.161 (0.086)           & \comb             & 0.160 (0.082)   & \mytheta            & \underline{0.188} (0.109)                        \\
			\textbf{3}                       & \mytheta            & 0.176 (0.105)            & \mytheta            & \underline{0.157} (0.042) & \holt & 0.166 (0.076)    & \mytheta           & 0.163 (0.087)           & \mytheta            & \underline{0.162} (0.086)   & \damped           & \underline{0.189} (0.110)                        \\
			\textbf{4}                       & \ses              & 0.177 (0.106)            & \ses              & \underline{0.158} (0.028) & \rf & 0.174 (0.066)   & \holt            & 0.188 (0.074)           & \ses              & \underline{0.163} (0.094)   & \comb             & 0.190 (0.112)                        \\
			\textbf{5}                       & \holt             & 0.179 (0.113)            & \holt             & 0.159 (0.036) & \ses & 0.183 (0.087)   & \ses             & 0.212 (0.070)           & \holt             & 0.171 (0.119)   & \holt             & 0.195 (0.118)                        \\
			\textbf{6} & \rf               & 0.232 (0.150)            & \rf               & 0.200 (0.025) & \damped & 0.212 (0.044)   & \rf              & 0.287 (0.083)           & \rf               & 0.210 (0.094)   & \rf               & 0.207 (0.137)  \\
			\hline
		\end{tabular}
	}
\end{table}

Overall, the obtained rankings are consistent with the ground truth, proving the ability of the method to carry out model selection, by identifying the model with the lowest error measure. Moreover, comparing our approach with the \baseline{}, we find that our framework largely outperforms the latter, in that the ranking resulting from the \baseline{} is very different from the true one. Even in predictions with a small forecasting horizon ($h=90$), the \baseline{}'s ranking performance remains sub-optimal . 
Looking at the four \monitoring{}, we find that they have a different behavior depending on the dataset. Specifically, GPs result to be slightly more reliable than Bayesian-LeNet, as the latter in some cases swapped the first and second model of the ranking. LSTM's performance is close to the two probabilistic models, although the latter two globally have a better performance in terms of RMSE (see  Table~\ref{tab:summary-travel}). 

\begin{figure}[t]
	\centering
	\includegraphics[width=\textwidth]{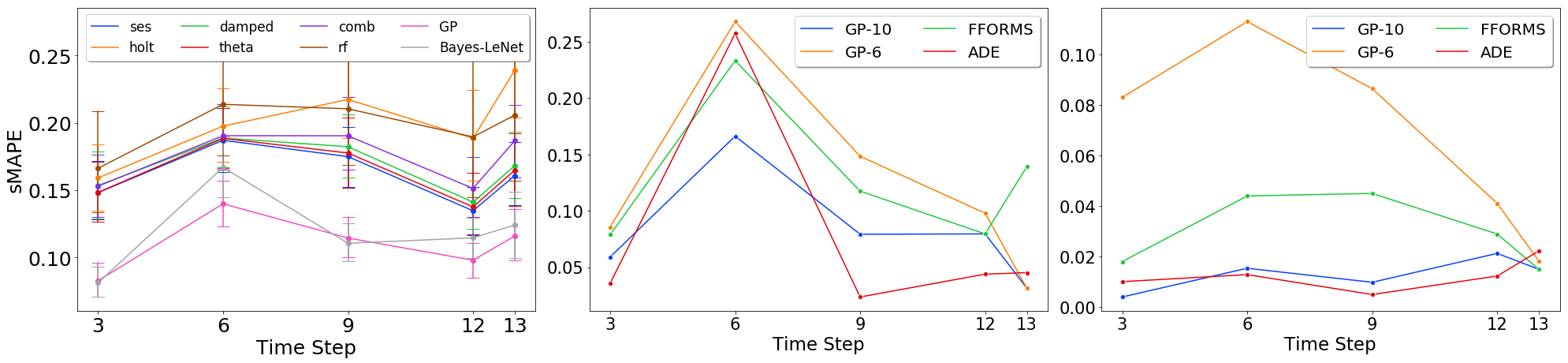}
%
	\caption{Measured average forecasting performance(sMAPE) using the proposed method for model selection in the \textsc{weekly} dataset with fixed forecasting models over the whole horizon. Average performance with Bayes-LeNet and GPs as \monitoring{} (left). Error bars denote standard deviation. Using GPs as \textit{monitoring model} with six (GP-6) and ten \monitored{} (GP-10), worst (center) and best (right) model selection performances in comparison with \tsensembler{} and \hyndmeta{}.}	\label{fig:naive_comp}
\end{figure}

Having showed the reliability of the rankings, we evaluate if these can be effectively used to maintain accurate forecasts over time by doing model selection at fixed periods of time. Specifically, given a forecasting horizon, we divide it in smaller periods. At each time point, we use the predicted forecasting error to rank the \monitored{} and thus perform model selection by picking the best ranked model. 
%
We use the public benchmark data to guarantee curated data and we limit the experiments to the best two \monitoring{}, Bayesian-LeNet and GPs (Table~\ref{tab:summary-travel}). We compare our model selection with the results obtained using the same \textit{monitored model} along the forecasting horizon. 
Figure \ref{fig:naive_comp} left shows the average forecasting performance, measured through the real sMAPE, on the \textsc{weekly} dataset. The proposed model selection scheme allows to have the lower forecasting errors, i.e. a better performance, along the whole forecasting horizon. Among the two \monitoring{}, GPs result in smoother curves. 

Finally, we compare with two state-of-the-art meta-learning methods, arbitrated dynamic ensembler~\cite{cerqueira2017arbitrated}, \tsensembler{}, and Feature-based FORecast-Model Selection~\cite{RePEc:msh:ebswps:2018-6}, \hyndmeta{}, with the best performing \textit{monitoring model} in our approach. The characteristics of these two methods allows them to be used to achieve good  forecasting model's performance. \hyndmeta{} uses 12 different base models, whereas \tsensembler{} uses up to 40 different models. To remain competitive with these two methods that use a larger number of base models, we add  three standard forecasting models, Arima (\arima), Random Walk (\rwf) and TBATS (\tbats)~\cite{de2011forecasting}, and  a feed-forward neural network (\nn), to our set of \monitored{}. We present sMAPE results over two time-series from the \textsc{weekly} dataset: one where our method performs worst (Fig.~\ref{fig:naive_comp} center) and the one where it performs best (Fig.~\ref{fig:naive_comp} right). We show the results of our approach using the original six \monitored{} and the enlarged set. Using the original six \monitored{}, our performance is worse than the two meta-learning models. However, by enlarging the set of \monitored{}, our method performs better than \hyndmeta{} and achieves a performance comparable to \tsensembler{} with much less monitored/base models.

\subsection{Model monitoring and selection performance}\label{sec:allperf}
Finally, we illustrate the performance of the proposed model monitoring and selection framework by using it to guarantee continuous price forecasting accuracy of our two travel products: flights and hotels. In this context, the predicted sMAPE is used as a surrogate measure of the quality of the forecasts estimated by the \monitored{}. When the predicted sMAPE surpasses a given threshold, model selection is performed. Otherwise, the \textit{monitored model} is kept. We use the best performing \textit{monitoring model}, GPs. Since this is a probabilistic method, in addition to having a high predicted sMAPE, we add the condition of having a low uncertainty in the prediction. In our experiments, we set the sMAPE threshold at 0.02 for flights and 0.01 for hotels. The uncertainty was set at 0.01 for both. For this experiment, we removed \rf{} from the \monitored{} pool as it is the method giving the poorest performance. It is important to remark that differently from other approaches removing a method from the \monitored{} pool simply requires to stop generating forecasts with the removed model. No re-training of the \monitoring{} is required.

\begin{figure}[t]
	\centering
	\includegraphics[width=0.7\textwidth]{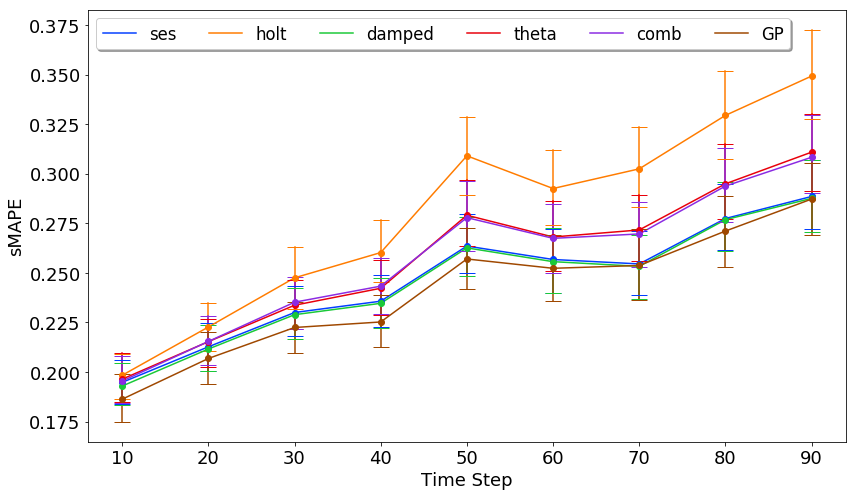} 
	\caption{Average forecasting performance in terms of sMAPE using the proposed model monitoring and selection framework (GPs as \textit{monitoring model}) and using forecasting fixed models over the whole horizon. Error bars denote the standard deviation.}
	\label{fig:final}
\end{figure}

Figure~\ref{fig:final} illustrates the results obtained in terms of the average performance (sMAPE) for \textsc{hotels} with forecasting horizon $h=90$. Our experiment here is quite restrictive, in the sense that no \textit{monitored model} is re-trained along the forecasting period. In this way, we show that even under this restrictive setting the proposed framework is able to improve the performance of simple models. This suggests that through the use of this framework it is possible to extend the moment where \monitored{} need to be re-trained by simply using the ranking information to pick a new model. Delaying model re-training represents important cost savings.
\section{Conclusions} \label{sect_conclusions}
In this paper we introduce a data-driven framework to constantly monitor and compare the performance of deployed time-series forecasting models to guarantee accurate forecasts of travel products' prices over time. The proposed approach predicts the forecasting error of a forecasting model and considers it as a surrogate of the model's future performance. The estimated forecasting error is hence used to detect accuracy deterioration over time, but also to compare the performance of different models and carry out dynamic model selection by simply ranking the different forecasting models based on the predicted error measure and selecting the best. In this work, we have chosen to use the sMAPE as forecasting performance measure, since it is appropriate for our application but, it cannot be used in settings where the time-series could present zero-valued observations. However, the framework is general enough that any other measure could be used instead.

The proposed framework has been designed to guarantee accurate price forecasts of different travel products price and it is conceived for travel applications that might be already deployed.  As such, it was undesirable to propose a method that performs forecasting and monitoring altogether, as in meta-learning, since this would require deprecating already deployed models to implement a new system. Instead, thanks to the proposed fully data-driven approach,  \textit{monitoring models} are completely independent of those doing the forecasts, i.e. the \monitored{}, thus allowing a transparent implementation of the monitoring and selection framework. 

Although our main objective is to guarantee stable accurate price forecasts, the problem we address is relevant beyond our concrete application. Sculley \textit{et al.}~\cite{Sculley2015} introduced the term hidden technical debt to formalize and help reason about the long term costs of maintainable ML systems. According to their terminology, the proposed model monitoring and selection framework addresses two problems: 1) the monitoring and testing of dynamic systems, which is the task of continuously assessing that a system is working as intended; and 2) the production management debt, which refers to the costs associated to the maintenance of a large number of models that run simultaneously. Our solution represents a simple, flexible and accurate alternative to these problems.


\end{document}